\documentclass[a4paper,12pt]{article}
\usepackage[utf8x]{inputenc}
\usepackage{amsmath}
\usepackage{amssymb}
\usepackage{graphicx}

\title{Solar and Atmospheric Neutrinos: Background Sources for the Direct Dark Matter Searches}
\author{A.~G\"utlein, C.~Ciemniak, F.~von~Feilitzsch, N.~Haag,\\ M.~Hofmann, C.~Isaila, T.~Lachenmaier, J.-C.~Lanfranchi,\\ L.~Oberauer, S.~Pfister, W.~Potzel, S.~Roth, M.~von~Sivers,\\ R.~Strau\ss{}, and A.~Z\"oller\\\\Physik-Department E15,\\ Technische Universit\"at M\"unchen, D-85748 Garching}

\date{29.03.2010}

\begin{document}
\maketitle

\begin{abstract}
In experiments for direct dark matter searches, neutrinos coherently scattering off nuclei can produce similar events as Weakly Interacting Massive Particles (WIMPs). The calculated count rate for solar neutrinos in such experiments is a few events per ton-year. This count rate strongly depends on the nuclear recoil energy threshold achieved  in the experiments for the WIMP search. We show that solar neutrinos can be a serious background source for direct dark matter search experiments using Ge, Ar, Xe and CaWO$_4$ as target materials. To reach sensitivities better than $\sim10^{-10}$\,pb for the elastic WIMP nucleon spin-independent cross section in the zero-background limit, energy thresholds for nuclear recoils should be $\gtrsim2.05$\,keV for CaWO$_4$, $\gtrsim4.91$\,keV for Ge, $\gtrsim2.89$\,keV for Xe, and $\gtrsim8.62$\,keV for Ar as target material. Next-generation experiments should not only strive for a reduction of the present energy thresholds but mainly focus on an increase of the target mass. Atmospheric neutrinos limit the achievable sensitivity for the background-free direct dark matter search to $\gtrsim10^{-12}$\,pb.
\end{abstract}

\section{Direct dark matter search}
There are many hints for the existence of non-baryonic cold dark matter \cite{Jungman96}, \cite{Clowe06}, \cite{Spergel07}, \cite{Bertone04}. Until now, however, it has not been possible to directly detect dark matter. At present there are several experiments searching for dark matter particles, in particular for WIMPs motivated by supersymmetric theories \cite{Jungman96}, \cite{Bertone04}. Due to the very low expected count rates these experiments need a highly effective background suppression. Most background events originate from the interaction of incident particles (e.g. gammas, betas) with the electrons of the target material (electron recoils), while WIMP interactions are expected to induce nuclear recoils in the target material. Most of the present detectors for direct dark matter searches can distinguish between electron recoils and nuclear recoils, see, e.g., ArDM \cite{ArDM09}, CDMS \cite{CDMS09}, CRESST \cite{CRESST09}, EDELWEISS \cite{Edelweiss09}, WARP \cite{WARP07}, XENON \cite{XENON08}, XENON100 \cite{XENON100}. In this way, such background events can be discriminated. Neutrinos, however, could mimic a WIMP event due to coherent neutrino nucleus scattering.

The present best limit for the WIMP-nucleon cross section is $\sim3\cdot10^{-8}$\,pb $\hat{=}3\cdot10^{-44}$\,cm$^2$ for a WIMP mass in the region of 50 - 70\,GeV \cite{CDMS09}, \cite{XENON100}. In \cite{CDMS09} this sensitivity was achieved with a raw exposure of 612 kg-days (germanium) showing 2 events which cannot be excluded as signal but which could also be background with 23\% probability\cite{CDMS09}. Thus, for the next-generation ton-scale experiments, in addition to the already challenging background sources like neutrons also solar neutrinos could become a dangerous background with a rate of several events per ton-year.

Coherent neutrino nucleus scattering as a background for dark matter searches was first suggested in \cite{Cabrera85} and \cite{Drukier86}, and has been calculated in \cite{Strigari09}, \cite{Monroe07}, \cite{Vergados08}. The calculations in the present paper extend beyond previously published work by considering calcium tungstate (CaWO$_4$) and sodium iodide (NaI) as target materials and include all parts of the solar neutrino spectrum (see sections \ref{countRate} and \ref{solarNeutrinos}). Some of our results are in disagreement with those reported in \cite{Monroe07} and \cite{Vergados08}.

\section{Coherent Neutrino Nucleus Scattering\\(CNNS)}
CNNS is a neutral-current interaction and thus independent of the neutrino flavour. A neutrino scatters off a nucleus by exchanging a virtual Z$^0$ boson. For low transferred momenta, the wavelength of the Z$^0$ boson is in the same order of magnitude as the radius of the nucleus. Thus, the neutrino scatters off all nucleons coherently leading to an enhanced cross section \cite{Drukier84}:
\begin{eqnarray}
\frac{d\sigma}{d\cos\theta} & = & \frac{G_F^2}{8\pi}\left[Z\left(4\sin^2\theta_W - 1\right) + N\right]^2 E_{\nu}^2\left(1 + \cos\theta\right) |f(q)|^2\\
\frac{d\sigma}{dE_{rec}} & = & \frac{G_F^2}{8\pi}\left[Z\left(4\sin^2\theta_W - 1\right) + N\right]^2 M \left(2-\frac{E_{rec}M}{E_{\nu}^2}\right) |f(q)|^2\\
\sigma_{tot} & = &  \frac{G_F^2}{4\pi}\left[Z\left(4\sin^2\theta_W - 1\right) + N\right]^2 E_{\nu}^2 |f(q)|^2\label{equ:sigmaTot}
\end{eqnarray}
where $G_F$ is the Fermi constant, $Z$ the proton number of the target nucleus, $N$ the neutron number of the target nucleus, $\theta_W$ the Weinberg angle, $E_{\nu}$ the neutrino energy, $\theta$ the scattering angle in the laboratory frame, $M$ the mass of the target nucleus and $E_{rec}$ its recoil energy. For larger transferred momenta $q=\sqrt{2ME_{rec}}$ the Z$^0$ wavelength is smaller than the target nucleus. Thus, the cross section is modified by a form factor $|f(q)|^2$. For this work we used the Helm form factor \cite{Helm56}:
\begin{equation}\label{equ:HelmFF}
	|f(q)|^2 = \left(3\frac{j_1(qR_0)}{qR_0}\right)^2e^{-q^2s^2}
\end{equation}
where $j_1$ is the spherical Bessel function of the first kind and order 1, $R_0 = 1.14 A^{\frac{1}{3}}$\,fm \cite{Lewin96} the effective radius of the target nucleus, and $s = 0.9$\,fm \cite{Lewin96} the skin thickness of the nucleus. For the calculation of this form factor the distribution of the charge is assumed to be a convolution of a constant inner part and a gaussian distribution modeling the skin of the nucleus. We used the Helm form factor to compare our results with those of \cite{Strigari09}.

Since $\sin^2\theta_W = 0.23$ \cite{PDG04}, the total cross section $\sigma_{tot}$ is approximately
\begin{equation}\label{equ:sigmaTotApporx}
\sigma_{tot} \approx \frac{G_F^2}{4\pi} N^2 E_{\nu}^2 |f(q)|^2 = 4.2\cdot 10^{-45} N^2\left(\frac{E_{\nu}}{1\text{\,MeV}}\right)^2 |f(q)|^2 \text{\,cm}^2.
\end{equation}

\section{Count rate calculation}\label{countRate}
\begin{figure}[htbp]
	\centering
	\includegraphics[width=0.8\textwidth]{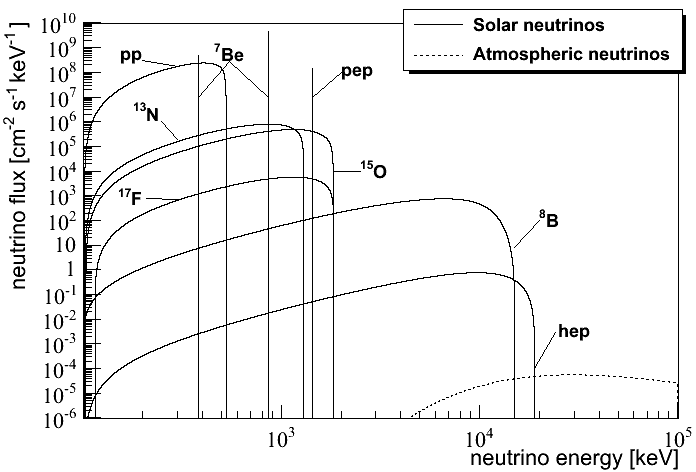}
	\caption{Energy spectra of solar (solid lines) and atmospheric (dashed lines) neutrinos (see section \ref{Atmospheric}).}
	\label{fig:NeutrinoSpectra}
\end{figure}
The spectra for solar \cite{Bahcall05}, \cite{BahcallHP} and atmospheric neutrinos \cite{Battistoni05} are depicted in figure~\ref{fig:NeutrinoSpectra}. The count rate for neutrinos is given by
\begin{equation}\label{equ:CountRate}
R = N_t \int_0^{\infty} dE_{\nu}\Phi(E_{\nu})\int_0^{\frac{2E_{\nu}^2}{M}}dE_{rec} \frac{d\sigma(E_{\nu}, E_{rec})}{dE_{rec}}
\end{equation}
where $N_t$ is the total number of target nuclei, $\Phi(E_{\nu})$ the neutrino flux, and $\frac{2E_{\nu}^2}{M}$ the maximal recoil energy (i.e., back scattering of the neutrino at $\theta=\pi$). For the simple case of monoenergetic solar pep neutrinos the flux is given by \cite{Bahcall05}:
\begin{equation}\label{equ:pepFlux}
\Phi(E_{\nu})_{pep} = \delta(E_{\nu} - 1.442\text{\,MeV})\cdot 1.41\cdot 10^{8} \frac{1}{\text{cm}^2\text{s}\text{MeV}}.
\end{equation}
In the case of monoenergetic pep neutrinos the second integral in equation (\ref{equ:CountRate}) can be calculated leading to the expression for the total cross section in equation (\ref{equ:sigmaTot}). With the approximation (\ref{equ:sigmaTotApporx}) for the total cross section, the count rate for pep neutrinos is
\begin{equation}\label{equ:PEPsimple}
R_{pep} = N_t\cdot 1.41\cdot 10^{8} \frac{1}{\text{cm}^2\text{s}} \cdot 4.2\cdot 10^{-45} N^2\cdot\left(1.442\text{\,} \frac{\text{MeV}}{1 \text{MeV}}\right)^2 \text{\,cm}^2,
\end{equation}
where $|f(q)|^2\approx1$ in good approximation (the corrections are a few percent only). For a germanium target, for example, pep neutrinos produce a count rate of $R_{pep}\sim539$ events per ton-year (for zero energy threshold, see below).

For the calculation of count rates and recoil spectra of other parts of the solar neutrino spectrum, the two integrals in equation (\ref{equ:CountRate}) can be swapped. Then, the integration limits have to be changed too (the lower limit is determined by the minimum neutrino energy $E_{\nu}$ leading to the recoil energy $E_{rec}$):
\begin{equation}\label{equ:CountRateNew}
R = N_t \int_0^{\infty} dE_{rec}\int_{\sqrt{\frac{E_{rec}M}{2}}}^{\infty}dE_{\nu}\Phi(E_{\nu})\frac{d\sigma(E_{\nu}, E_{rec})}{dE_{rec}}.
\end{equation}
The recoil spectrum $\frac{dR\left(E_{rec}\right)}{dE_{rec}}$ is given by the second integral in equation (\ref{equ:CountRateNew})
\begin{equation}
\frac{dR\left(E_{rec}\right)}{dE_{rec}} = N_t \int_{\sqrt{\frac{E_{rec}M} {2}}}^{\infty}dE_{\nu}\Phi(E_{\nu})\frac{d\sigma(E_{\nu}, E_{rec})}{dE_{rec}}. 
\end{equation}

In order to evaluate the number of events due to CNNS in a detector, the number of events above a given energy threshold $E_{th}$ is calculated. It is important to state what energy threshold means in this context. The energy threshold which is applied in the following calculations is the minimal nuclear recoil energy used for the dark matter search. The count rate above an energy threshold $E_{th}$ is given by:
\begin{equation}
R_{th} =\int_{E_{th}}^{\infty} dE_{rec} \frac{dR\left(E_{rec}\right)}{dE_{rec}} = N_t \int_{E_{th}}^{\infty} dE_{rec}\int_{\sqrt{\frac{E_{rec}M}{2}}}^{\infty}dE_{\nu}\Phi(E_{\nu})\frac{d\sigma(E_{\nu}, E_{rec})}{dE_{rec}}.
\end{equation}

\section{Count rates for solar neutrinos}\label{solarNeutrinos}
The solar neutrino fluxes are taken from \cite{Bahcall05} and the spectra (see figure~\ref{fig:NeutrinoSpectra}) of the different solar neutrinos from \cite{BahcallHP}.
\begin{figure}[htbp]
	\centering
	\includegraphics[width = 0.425\textwidth]{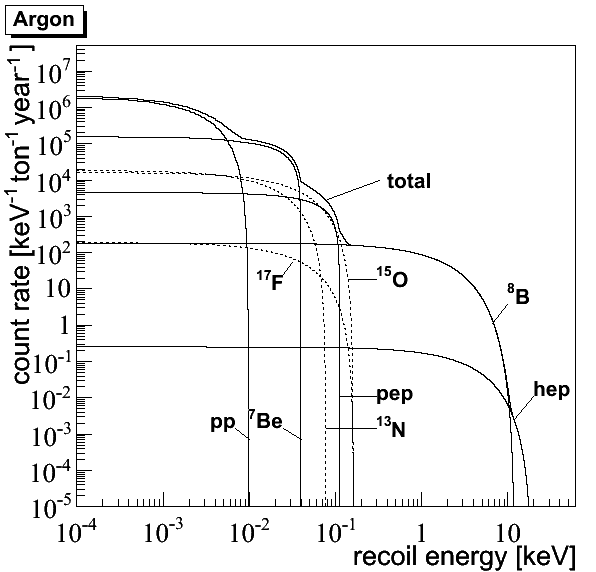}
	\includegraphics[width = 0.425\textwidth]{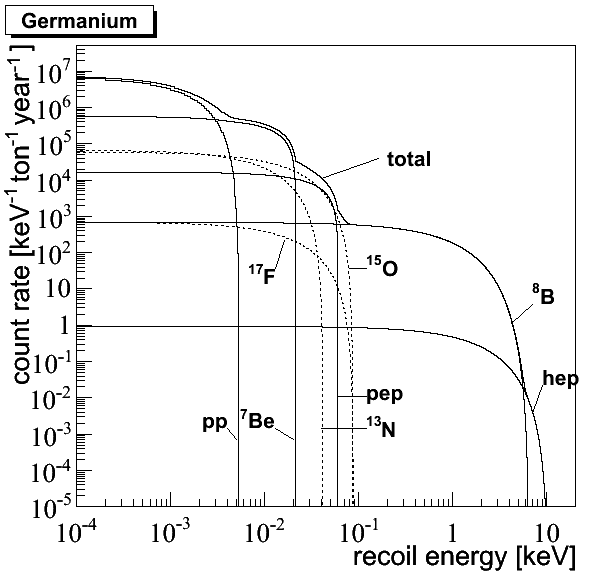}
	\includegraphics[width = 0.425\textwidth]{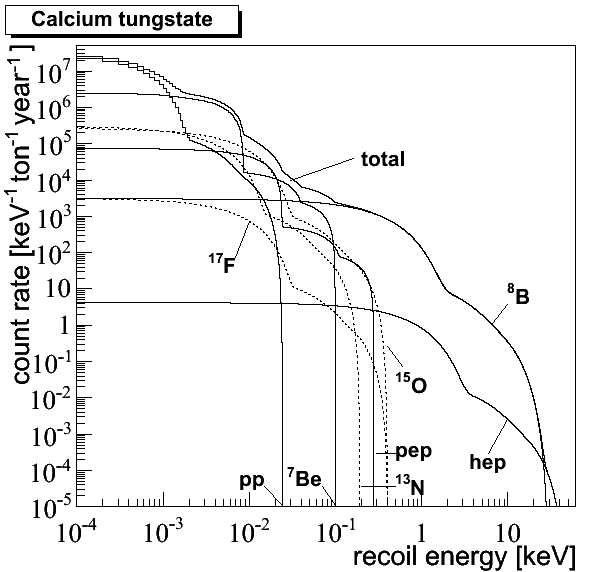}
	\includegraphics[width = 0.425\textwidth]{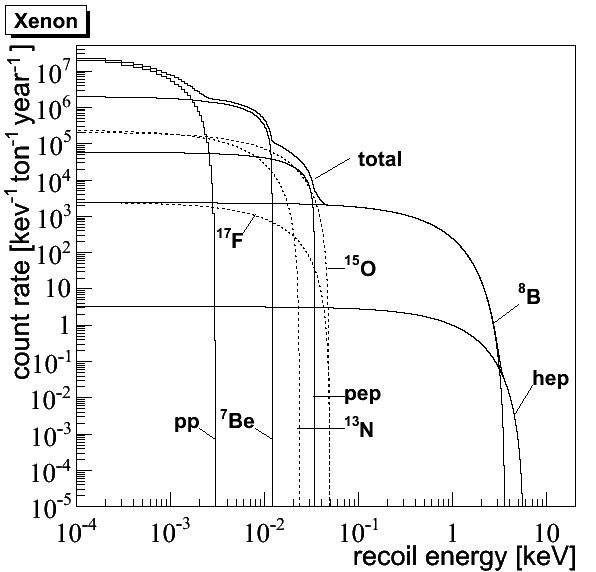}
	\includegraphics[width = 0.425\textwidth]{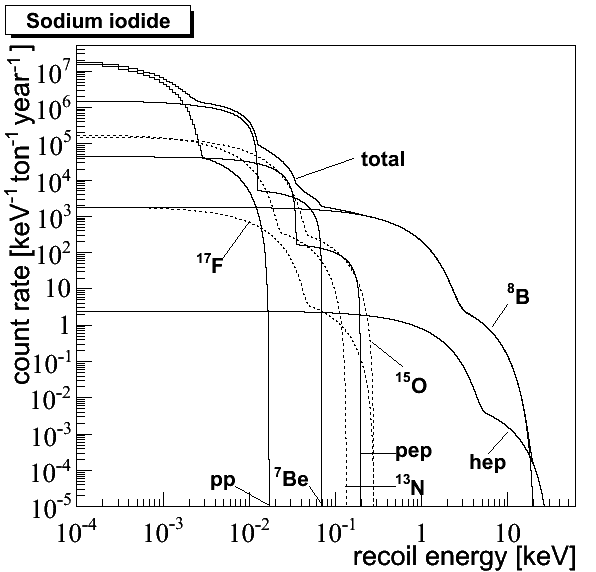}
	\caption{Nuclear recoil spectra from solar neutrino scattering for different target materials. For the energy thresholds of the present dark matter experiments only $^{8}$B and hep neutrinos contribute to the count rate. The dashed lines indicate the contribution of the CNO cycle to the neutrino flux. The kinks in the spectra for CaWO$_4$ and NaI arise from the different contributions of the elements in these materials.}
	\label{fig:Solar}
\end{figure}
Figure~\ref{fig:Solar} depicts the recoil spectra of solar neutrinos for different target materials.
\begin{figure}[htbp]
	\centering
	\includegraphics[width = 0.425\textwidth]{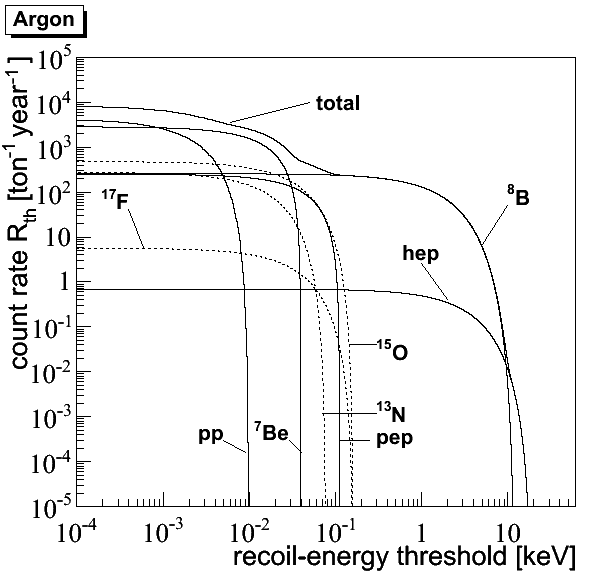}
	\includegraphics[width = 0.425\textwidth]{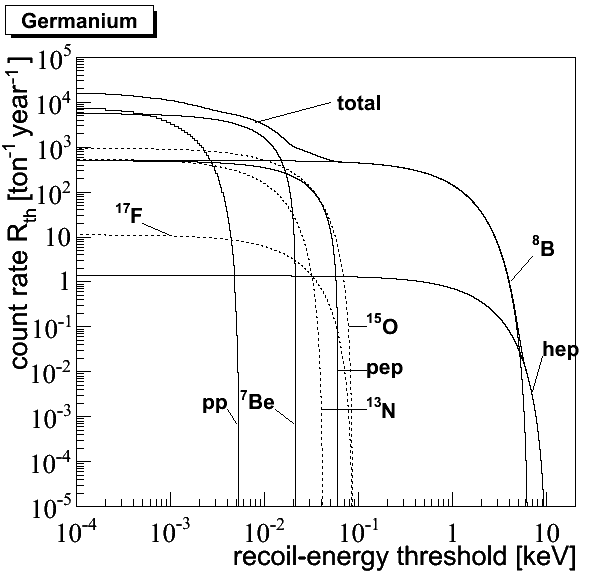}
	\includegraphics[width = 0.425\textwidth]{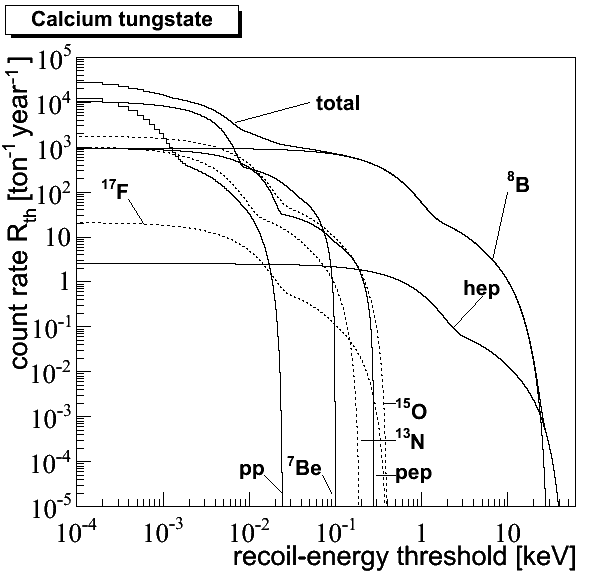}
	\includegraphics[width = 0.425\textwidth]{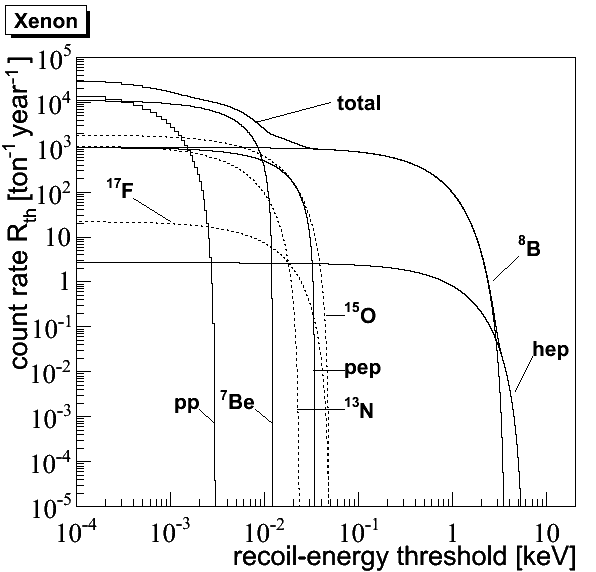}
	\includegraphics[width = 0.425\textwidth]{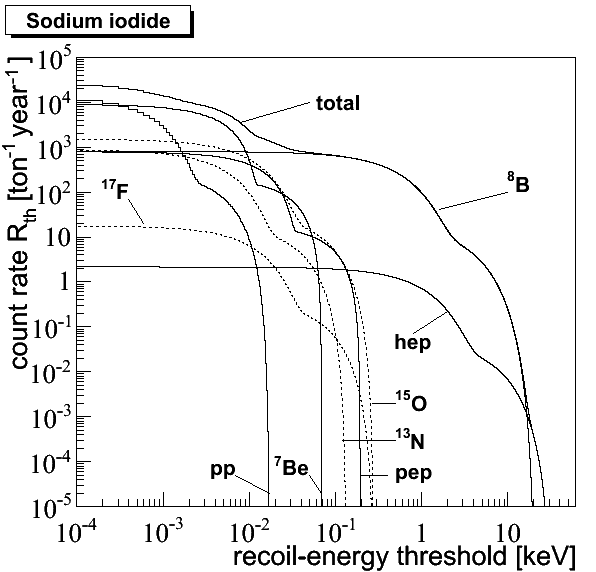}
	\caption{Integrated count rates from solar neutrino scattering above recoil-energy threshold for different target materials. For other details, see caption of figure \ref{fig:Solar}.}
	\label{fig:SolarInt}
\end{figure}
In figure~\ref{fig:SolarInt}, the integrated count rates above the recoil-energy threshold (horizontal axis) are plotted.
\begin{table}[htbp]
	\centering
	\begin{tabular}{|l|r|r|r|r|r|r|}
		\hline
		\textbf{Material} & \textbf{0\,keV} & \textbf{1\,keV} & \textbf{2\,keV} & \textbf{3\,keV} & \textbf{5\,keV} & \textbf{10\,keV}\\
		\hline
		Ar & 8.0$\cdot10^{3}$ & 1.3$\cdot10^{2}$ & 6.5$\cdot10^{1}$ & 3.2$\cdot10^{1}$ & 6.2 & 1.5$\cdot10^{-2}$ \\
		\hline
		Ge & 1.6$\cdot10^{4}$ & 1.4$\cdot10^{2}$ & 3.6$\cdot10^{1}$ & 7.2 & 7.9$\cdot10^{-2}$ & 2.5$\cdot10^{-8}$\\
		\hline
		NaI & 2.4$\cdot10^{4}$ & 9.6$\cdot10^{1}$ & 1.4$\cdot10^1$ & 6.3 & 2.7 & 2.4$\cdot10^{-1}$ \\
		\hline
		Xe & 3.0$\cdot10^{4}$ & 9.5$\cdot10^{1}$ & 4.7 & 6.2$\cdot10^{-2}$ & 9.0$\cdot10^{-5}$ & - \\
		\hline
		ZnWO$_4$ & 2.7$\cdot10^{4}$ & 6.6$\cdot10^{1}$ & 1.8$\cdot10^{1}$ & 9.7 & 4.4 & 1.0\\
		\hline
		CaWO$_4$ & 2.8$\cdot10^{4}$ & 5.8$\cdot10^{1}$ & 1.8$\cdot10^{1}$ & 1.2$\cdot10^{1}$ & 5.4 & 1.1 \\
		\hline
		\hline
		O & 5.8$\cdot10^{2}$ & 1.4$\cdot10^{1}$ & 1.1$\cdot10^{1}$ & 8.1 & 4.6 & 1.1 \\
		\hline
		Ca & 9.0$\cdot10^{2}$ & 1.4$\cdot10^{1}$ & 7.2 & 3.5 & 6.9$\cdot10^{-1}$ & 1.7$\cdot10^{-3}$ \\
		\hline
		W & 2.6$\cdot10^{4}$ & 2.9$\cdot10^{1}$ & 1.4$\cdot10^{-1}$ & 2.3$\cdot10^{-3}$ & - & - \\
		\hline
	\end{tabular}
	\caption{Nuclear recoil count rates from solar neutrino scattering per ton-year for various target materials and different energy thresholds. The count rates for the different nuclei of calcium tungstate are given separately.}
	\label{tab:SolarCOuntRates}
\end{table}
In table~\ref{tab:SolarCOuntRates} the count rates for various target materials and different energy thresholds are summarized. These count rates are calculated per ton of target material and year of exposure for an energy window between the minimum energy thresholds as given in table \ref{tab:SolarCOuntRates} and a maximum energy threshold of 100\,keV.

For example, the current energy threshold for xenon is 4.5\,keV \cite{XENON08} leading to $1.16\cdot10^{-3}$ events per ton-year. For germanium and calcium tungstate (CaWO$_4$) the energy threshold used is $\sim10$\,keV \cite{CDMS09}, \cite{CRESST09} producing no events ($\sim10^{-8}$) in germanium and $\sim1$ event per ton-year in calcium tungstate mainly due to oxygen recoils. All these events are only due to solar $^8$B and hep neutrinos. The other parts of the solar neutrino spectrum become important only for energy thresholds below 1\,keV.

The results of table~\ref{tab:SolarCOuntRates} are in good agreement with the results of \cite{Strigari09} for germanium, xenon and argon and they are also comparable to the simple approximation applied to the pep neutrino count rate in equation (\ref{equ:PEPsimple}). However, our results are lower than those obtained in \cite{Monroe07} by a factor of $\sim3$ and are higher by a factor of up to $\sim10^{3}$ than those of \cite{Vergados08}. Since the changes due to the form factor $|f(q)|^2$ (see equation (\ref{equ:HelmFF})) are only a few percent, the disagreement with the results of \cite{Monroe07} and \cite{Vergados08} must have a different origin. In the present paper we include also results for calcium tungstate and sodium iodide.

\section{Discussion}
\subsection{Solar neutrinos}
\begin{table}[htbp]
	\centering
	\begin{tabular}{|l|r|}
		\hline
		\textbf{Nucleus} & \textbf{Max. recoil energy} \\
		\hline
		O & 47.1\,keV\\
		\hline
		Na & 37.6\,keV\\
		\hline
		Ar & 18.8\,keV\\
		\hline
		Ca & 18.8\,keV\\
		\hline
		Zn & 11.6\,keV\\
		\hline
		Ge & 10.3\,keV \\
		\hline
		I & 5.9\,keV\\
		\hline
		Xe & 5.7\,keV\\
		\hline
		W & 4.1\,keV\\
		\hline
	\end{tabular}
	\caption{Maximum nuclear recoil energies for different target materials for an energy of solar neutrinos of 18.79\,MeV (i.e. the maximum hep neutrino energy).}
	\label{tab:MaximumRecoil}
\end{table}
In table~\ref{tab:MaximumRecoil} the maximum  nuclear recoil energies due to solar neutrinos for different target nuclei are summarized. For energy thresholds above these maximum recoil energies, solar neutrinos are not a background source for direct dark matter searches.

\subsubsection{Calcium tungstate (CaWO$_4$) (CRESST)}\label{CaWO4}
\begin{figure}[htbp]
	\centering
	\includegraphics[width=0.8\textwidth]{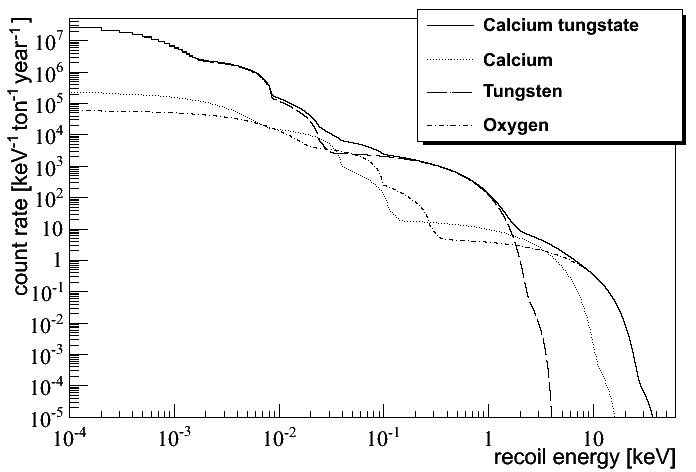}
	\caption{Recoil spectra of solar neutrinos for the different nuclei of calcium tungstate. The maximum recoil energy of tungsten is 4.09\,keV (see table~\ref{tab:MaximumRecoil}). Thus, for recoil energies above $\sim4$\,keV neutrinos only scatter off calcium or oxygen.} 
	\label{fig:Discrimination}
\end{figure}
Figure~\ref{fig:Discrimination} shows the recoil spectra of solar neutrinos for the different nuclei of calcium tungstate. The maximum recoil energy (i.e. backscattering of the neutrino) is 4.1\,keV for tungsten as target nucleus. Thus, the events above an energy threshold of $\sim4$\,keV are only due to scattering off calcium or oxygen. If a calcium tungstate detector is able to distinguish between tungsten recoils (i.e., the incident particle scatters off tungsten as expected for WIMPs) and calcium or oxygen recoils \cite{Lanfranchi08}, only tungsten recoils can be used for the WIMP search and the solar neutrino events above 4\,keV can be rejected. However, without this discrimination solar neutrinos lead to $\sim1$ event per ton-year for the present energy threshold of $\sim10$\,keV. Thus, the discrimination between tungsten recoils and recoils of oxygen or calcium is crucial for the use of calcium tungstate in future ton-scale experiments.

In order to perform a background-free measurement the count rates of solar neutrinos should be 0.1 counts per ton-year, so that the probability for no events is 90.5\% due to the Poisson distribution. For a given WIMP mass of $\sim60$\,GeV a background-free experiment with an exposure of one ton-year can reject all WIMP-nucleon cross sections that predict a count rate higher than 2.3 counts per ton-year. To estimate the number of WIMP events, here and in the following examples we have assumed spin-independent elastic WIMP-nucleon scattering \cite{Lewin96} and a WIMP mass of $\sim60$\,GeV.

If the discrimination between tungsten recoils and oxygen or calcium recoils can be achieved \cite{Lanfranchi08} the energy threshold for calcium tungstate should be above 2.05\,keV so that solar neutrinos are leading to $\sim0.1$ tungsten recoil events per ton-year. With this energy threshold a background-free experiment using calcium tungstate as target material can reach a sensitivity of $\sim1.1\cdot10^{-11}$\,pb for the WIMP-nucleon cross section with a confidence level of 90\%.

Without the discriminetion between tungsten recoils and oxygen or calcium recoils the energy threshold should be 16.31\,keV to achieve 0.1 counts per ton-year for solar neutrinos. With this energy threshold the reachable sensitivity for the WIMP-nucleon cross section is $\sim6.8\cdot10^{-11}$\,pb.

The same conclusion we reach for zinc tungstate (ZnWO$_4$) which is pre\-sent\-ly also investigated as target material for direct dark matter searches \cite{ZINC}.

\subsubsection{Germanium (CDMS, EDELWEISS)}\label{Ge}
The present energy threshold for dark matter experiments with germanium is $\sim10$\,keV \cite{CDMS09} which is in agreement with the maximum recoil energy of solar neutrinos (see table~\ref{tab:MaximumRecoil}). For recoil energies below 4.91\,keV solar neutrinos are leading to $\sim0.1$ events per ton-year. For this energy threshold a background-free experiment with germanium as target material can achieve a sensitivity of $\sim1,7\cdot10^{-11}$\,pb for the WIMP-nucleon cross section with a confidence level of 90\%.

The calculated solar neutrino count rates for germanium are in agreement with \cite{Strigari09}. However, these count rates are smaller by a factor of $\sim3$ than those in \cite{Monroe07}.

\subsubsection{Xenon (XENON)}\label{Xe}
For the present energy threshold (4.5\,keV) of dark matter experiments using xenon as target material solar neutrinos would produce about 1.2$\cdot10^{-3}$ events per ton-year. Already for the slightly lower energy threshold of 2.89\,keV, solar neutrinos are leading to $\sim0.1$ events per ton-year. For this energy threshold experiments using xenon as target material can reach a sensitivity of $\sim8.6\cdot10^{-12}$\,pb for the WIMP-nucleon cross section with a confidence level of 90\%.

Our count rates for xenon are in agreement with \cite{Strigari09}, but our count rates are smaller by a factor of $\sim3$ than those of \cite{Monroe07} and higher by a factor of $\sim1000$ than those of \cite{Vergados08}.

\subsubsection{Argon (ArDM, WARP)}\label{Ar}
For argon as target material a recoil energy threshold of 8.62\,keV would lead to $\sim0.1$ solar neutrino events per ton-year. For this energy threshold experiments using argon can reach a sensitivity of $\sim5.0\cdot10^{-11}$\,pb for the WIMP-nucleon cross section with a confidence level of 90\%. The present recoil energy threshold of these experiments is between 30\,keV and 50\,keV \cite{ArDM09}, \cite{WARP07}, well above 8.62\,keV.

The calculated solar neutrino count rates for argon are in agreement with \cite{Strigari09} but our count rates are smaller by a factor of $\sim3$ than those of \cite{Monroe07}.

\subsection{Atmospheric neutrinos}\label{Atmospheric}
As shown in the previous section, the solar-neutrino background will not become a problem, if the energy thresholds applied for the dark-matter search (see section~\ref{countRate}) are selected with care. However, atmospheric neutrinos have higher energies leading to higher recoil energies.
\begin{figure}[htbp]
	\centering
	\includegraphics[width = 0.425\textwidth]{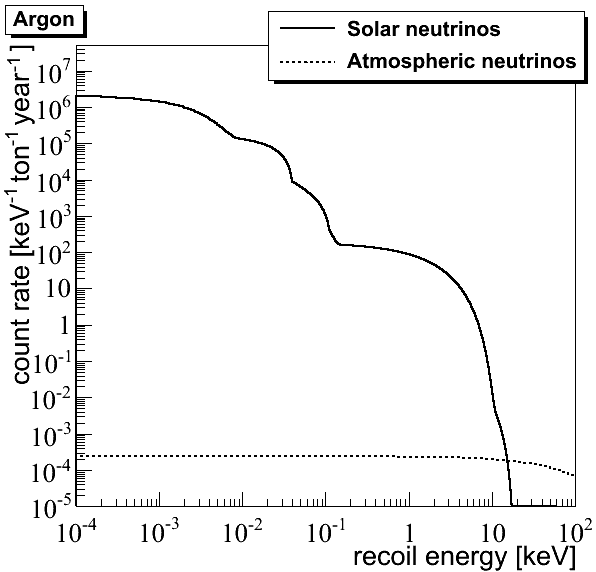}
	\includegraphics[width = 0.425\textwidth]{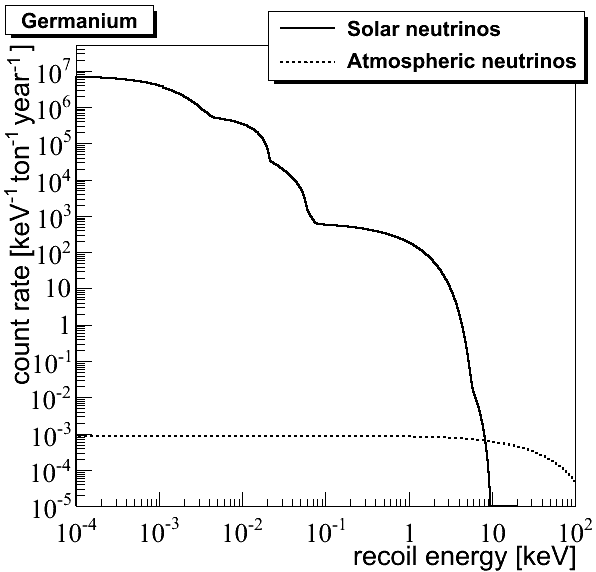}
	\includegraphics[width = 0.425\textwidth]{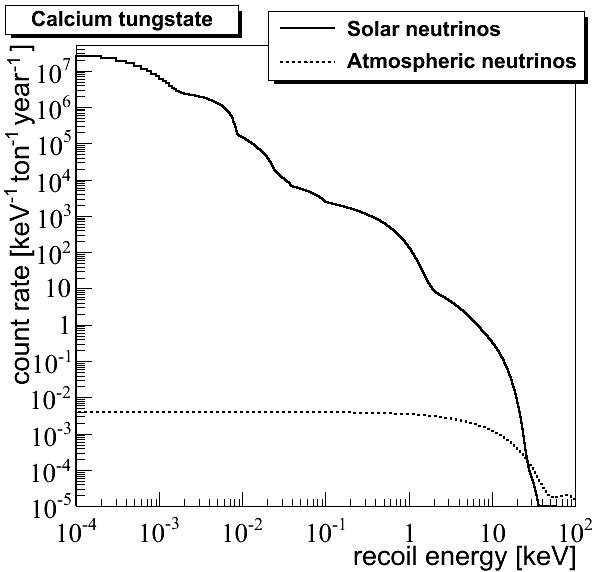}
	\includegraphics[width = 0.425\textwidth]{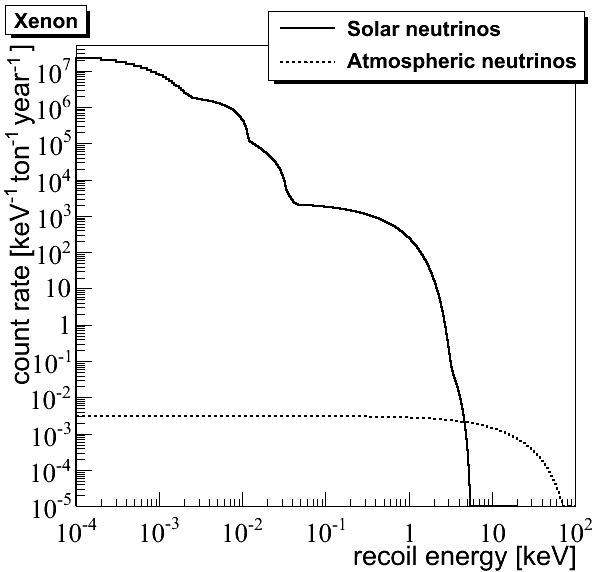}
	\includegraphics[width = 0.425\textwidth]{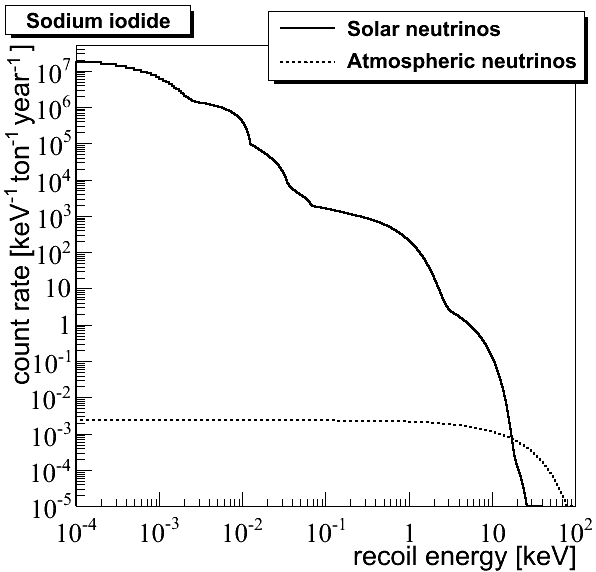}
	\caption{Recoil spectra of atmospheric ($\nu_{\mu}$, $\overline{\nu_{\mu}}$, $\nu_e$, $\overline{\nu_e}$) and solar neutrinos for different target materials. The feature in the recoil spectrum of atmospheric neutrinos for CaWO$_4$ at a recoil energy of $\sim50$\,keV is due to the form factor (see equation (\ref{equ:HelmFF})).}
	\label{fig:Atmospheric}
\end{figure}
Figure \ref{fig:Atmospheric} shows, for comparison, the recoil spectra of atmospheric and  solar neutrinos for different target materials. The spectra of atmospheric neutrinos were calculated in an analogous way as for solar neutrinos described in section \ref{countRate}. The flux and energy spectra ($\lesssim100$\,MeV) of atmospheric neutrinos ($\nu_{\mu}$, $\overline{\nu_{\mu}}$, $\nu_e$, $\overline{\nu_e}$) were taken from \cite{Battistoni05}. For neutrino energies above 100\,MeV the form factor becomes very small. Thus, for this calculation we took only atmospheric neutrinos with energies below 100\,MeV into account.

At energies $\lesssim100$\,MeV, the flux of atmospheric neutrinos has an uncertainty of $\sim20\%$ \cite{Strigari09}. The spectra in figure \ref{fig:Atmospheric} were calculated with the Helm form factor of equation (\ref{equ:HelmFF}). For the present energy thresholds of the dark matter experiments, atmospheric neutrinos lead to $\sim10^{-2}$ events per ton-year \cite{Strigari09}, \cite{Monroe07}. Our results are in agreement with those of \cite{Strigari09}. In \cite{Monroe07} only calculations for carbon have been included. They derive a value which is a factor of $\sim100$ smaller. This is due to the much lower number of neutrons in carbon as compared to xenon.

As a consequence, atmospheric neutrinos will be a serious background for WIMP sensitivities better than $\sim10^{-12}$\,pb. This sensitivity can be reached with an exposure of $\sim10$ ton-years assuming the energy thresholds mentoined in sections \ref{CaWO4} - \ref{Ar}. To acquire even better sensitivities the recoil spectra of atmospheric and solar neutrinos would have to be fitted to the measured data. Assuming that for a reasonable fit at least $\sim100$ events due to neutrinos are necessary, an exposure of a few $10^{3}$ ton-years would be required to reach enough statistical data for a spectral fit. To achieve such high statistics is very challenging. Thus, if the next generation of dark-matter experiments fails to find WIMPs, the direct detection of WIMPs might become impossible with the present detection techniques and ideas. The DAMA/LIBRA collaboration has been taking a different - in this respect more favourable - route by looking for a characteristic WIMP signature, an annular modulation \cite{DAMA10} using a much larger target mass than other direct dark matter experiments.

\section{Conclusions}
Naturally occurring neutrinos are a dangerous background for direct dark matter searches. To reach sensitivities better than $\sim10^{-10}$\,pb for the WIMP-nucleon cross section, the solar neutrino events have to be rejected by cuts with nuclear recoil-energy thresholds of $\gtrsim2$\,keV for calcium tungstate (if tungsten recoils can be singled out, $\gtrsim16$\,keV otherwise), $\gtrsim5$\,keV for germanium, $\gtrsim3$\,keV for xenon, and $\gtrsim9$\,keV for argon as target material. Thus, next-generation experiments like EURECA \cite{EURECA07}, SuperCDMS \cite{SuperCDMS05}, ArDM \cite{ArDM09} and XENON1T \cite{XENON1T} should not only focus on a reduction of the present energy thresholds but mainly on the production of large absorber masses.

Atmospheric neutrinos become a dangerous background for sensitivities better than $\sim10^{-12}$\,pb for the WIMP-nucleon cross section. Atmospheric neutrinos can not be rejected because it is not possible to select a proper recoil-energy window where their contribution can be neglected without a drastic reduction of the sensitivity for the WIMP search. Thus, for sensitivities better than $10^{-12}$\,pb direct dark matter searches would suffer from neutrinos mimicking WIMPs. To reach better sensitivities it would be necessary to fit the recoil spectra of solar and atmospheric neutrinos to the measured data. Such a spectral fit would need high statistics (we assume at least about 100 events, see section \ref{Atmospheric}) and thus an exposure of a few $10^{3}$ ton-years. To reach such high exposures with the present techniques in direct dark matter experiments seems too demanding. However, detectors with, e.g., \textit{directional sensitivity} (see, e.g., \cite{Tilman10}) could reject solar neutrinos also for low-energy thresholds. Applying such a technique the sensitivity for the WIMP-nucleon interaction could be increased for the same exposure. A promising approach is the confirmation of an annular modulation of the WIMP intensity due to the motion of the earth relative to the WIMP stream \cite{DAMA10}. The DAMA/LIBRA collaboration observes a modulation with the same phase as the expected modulation of the WIMP intensity. According to our calculations, with an exposure of 1.17 ton-years \cite{DAMA10}, there should be $\sim15$ solar neutrino events (if we assume 100\% efficiency for the detection of solar neutrinos and a recoil energy between 2 and 6\,keV without quenching) within the complete measuring period which corresponds to $\sim10^{-5}$ events per keV, kg, and day. The amplitude of the measured modulation is $\sim10^{-2}$ events per keV, kg and day. In addition, and even more important, the phase of the expected modulation of the solar neutrino flux, due to the elliptic orbit of the earth around the sun, is shifted by about half a year compared to the modulation observed by DAMA/LIBRA. Thus, neutrinos can not be the reason for the measured modulation.

\section{Acknowledgment}
This work was supported by funds of the Deutsche Forschungsgemeinschaft DFG (Transregio 27: Neutrinos and Beyond), the Munich Cluster of Excellence (Origin and Structure of the Universe), and the Maier-Leibnitz-Laboratorium (Garching).

\end{document}